\documentclass[aps,prmaterials,reprint,superscriptaddress]{revtex4-2}

\usepackage{amssymb}
\usepackage{graphicx}

\usepackage[pdfusetitle,
 bookmarks=true,bookmarksnumbered=false,bookmarksopen=false,
 breaklinks=false,pdfborder={0 0 0},pdfborderstyle={},backref=false,colorlinks=true]
{hyperref}

\begin{document}
\title{Superconductivity on ScH$_{3}$ and YH$_{3}$ hydrides: Effects of applied pressure in combination with electron- and hole-doping on the electron-phonon coupling properties}
\author{S. Villa-Cort{\'e}s}
\email{svilla@ifuap.buap.mx}

\author{O. De la Pe{\~n}a-Seaman}
\affiliation{Instituto de F{\'i}sica, Benem{\'e}rita Universidad Aut{\'o}noma de Puebla, Apartado Postal J-48, 72570, Puebla, Puebla, M{\'e}xico }

\date{\today}

\begin{abstract}
The implementation of electron- and hole-doping, in conjunction to applied pressure, is analyzed as a mechanism to induce or enhance the superconducting state on fcc YH$_3$ and ScH$_3$. In particular, the evolution of their structural, electronic, and lattice dynamical properties, as well as the electron-phonon coupling and superconducting critical temperature ($T_c$) is presented 
and discussed, as a function of the electron- and hole-doping content as well as applied pressure. The study was performed within the density functional perturbation theory, taking into account the effects of zero-point energy through the quasi-harmonic approximation, while the doping was implemented by means of the construction of the Sc$_{1-x}$M$_{x}$H$_{3}$ (M=Ca,Ti) and Y$_{1-x}$M$_{x}$H$_{3}$ (M=Sr,Zr) solid solutions modeled with the virtual crystal approximation (VCA). 
We found that the ScH$_3$ and YH$_3$ hydrides shown a significant improvement of their electron-phonon coupling properties under hole-doping (M=Ca,Sr) and at pressure values close to dynamical instabilities. Instead, by electron-doping (M=Ti,Zr), the systems do not improve such properties, whatever value of applied pressure is considered. Then, as a result, $T_c$ rapidly increases as a function of $x$ on the 
hole-doping region, reaching its maximum value of $92.7(67.9)$~K and $84.5(60.2)$~K at $x=0.3$ for Sc$_{1-x}$Ca$_{x}$H$_{3}$ at $10.8$~GPa and Y$_{1-x}$Sr$_{x}$H$_{3}$ at $5.8$~GPa respectively, 
with $\mu^{*}=0(0.15)$, while for both, electron- and hole-doping, $T_c$ decreases as a function of the applied pressure, mainly due to phonon hardening. By the thorough analysis of the electron-phonon properties as a function of doping and pressure, we can conclude that the tuning of the lattice dynamics is a promising path for improving the superconductivity on both systems.
\end{abstract}

\pacs{33.15.Ta}
\keywords{superconductivity}

\maketitle

\section{Introduction}

Materials with a high-superconducting critical temperature ($T_c$) are of great interest since at ambient conditions could have many technological applications. The theoretical breakthrough and progress for reaching high-$T_c$ superconductivity came after the Ashcroft\cite{PhysRevLett.92.187002} pioneering idea, suggesting that hydrogen-rich materials could be promising candidates for high-$T_c$ superconductivity. After that, several theoretical predictions have been proposed and performed on the crystal structure, at high pressure, of stoichiometric and hydrogen-rich materials, for which their electronic structure, lattice dynamics, and electron-phonon (el-ph) coupling properties have been calculated\cite{cor6,Duan2019,PhysRevB.96.100502,BI2019}. As a result of those predictions, several metal hydrides were proposed as conventional-superconductor candidates with a $T_c$ near room-temperature\cite{10.1038/srep06968,Liu6990,Wang6463}. The experimental breakthrough came with the discovery of phonon-mediated superconductivity on H$_{3}$S, with a $T_{c}$ of $203$~K under pressures as high as $155$~GPa\cite{203,eina}. Some years latter, high-$T_c$ superconductivity measurements were reported in other compounds, all of them at high applied pressure values, like LaH$_{10}$ with $T_c$ in the range of $250$-$260$~K\cite{La50,PhysRevLett.122.027001} at $170$~GPa, and more recently, YH$_{9}$ with $T_c=262$~K at $182$~GPa\cite{PhysRevLett.126.117003} and YH$_6$ giving $T_c=220$~K at $183$~GPa\cite{YH6em}, as well as a top reported $T_c$ value of $287$~K in a carbonaceous sulfur-hydride at $267$~GPa\cite{maxtc}.

As a result of the available theoretical and experimental reports, it has been determined that the tendency for superconductivity depends upon the species used to build up the metal hydride (together with hydrogen), a high density of states at the Fermi level, and that the resulting hydride compound must have a large electron-phonon coupling related to the hydrogen atoms. In particular, some of the highest $T_c$ values have being obtained from hydrides constructed with elements that belong to the alkaline family as well as the scandium group (first group of the transition metals) \cite{BI2019}. Regarding this group, from calculations on YH$_{n}$, the $T_{c}$ was estimated around $305$-$326$~K at an applied pressure of $250$~GPa for $n=10$ \cite{Liu6990,BI2019}; while for $n=6$, $T_c$ was on the range of $251$-$264$~K at $120$~GPa\cite{BI2019}; and for $n=4$, lower $T_c$ values (around $84$-$95$~K) at $120$~GPa were reported \cite{Yinwei}. For this family, only the YH$_{9}$ and YH$_{6}$ have been studied experimentally, as already mentioned. For the ScH$_{n}$ family, $T_{c}$ in the range of $120$-$169$~K was predicted for different members, with $n=6,7,9,10$ and $12$, for an applied pressure above $250$~GPa\cite{Ye2018HighHO,BI2019}. All the already mentioned metal hydrides can be considered as chemically precompressed phases relative to pure hydrogen, where high pressure is necessary for metallization\cite{PhysRevLett.92.187002}.

\begin{figure}
\includegraphics[width=6.0cm]{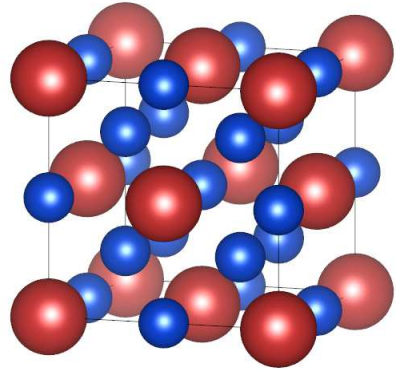}
\caption{\label{fig:crystalstructure} Fcc (cubic NaCl (B1)) structure (space group $Fm\bar{3}m$) of the Sc$_{1-x}$M$_{x}$H$_{3}$ and Y$_{1-x}$M$_{x}$H$_{3}$ solid solutions. The Scandium(Yttrium) and Hydrogen atoms are represented by large red and small blue spheres, respectively.}
\end{figure}

With respect to ScH$_{3}$ and YH$_{3}$ (metal-hydrides with low hydrogen content), they have hcp structures at ambient pressure, and are driven to fcc (cubic NaCl (B1) structure) (see Fig.\ref{fig:crystalstructure}) phase under applied pressure. 
For YH$_{3}$, Raman \cite{PhysRevB.76.024107} and infrared \cite{PhysRevB.73.104105} studies found that the cubic structure can exist at approximately $10$~GPa, and is clearly stabilized around $25$~GPa. It has been suggested that another intermediate phase \cite{PhysRevB.76.052101} or a coexisting hcp-cubic phase \cite{PALASYUK2005477} could also appear in the  $8$–$25$~GPa pressure range. 
It has been shown experimentally that YH$_{3}$ can be stabilized in the fcc phase at ambient pressure by substituting Y for 10$\%$ Li (Li$_{0.1}$Y$_{0.9}$H$_{3}$)\cite{LiYH31,LiYH3}. Recently, J. Purans \textit{et al.}\cite{YH3em} and Kong \textit{et al.}\cite{YH6em} reported the syntheses of this metal-hydride with a pure metallic fcc phase at a broad pressure interval, $40-340$~GPa. For lower applied pressure values, between $20$ and $40$~GPa, they found it to be a semi-metal with a distorted fcc crystal structure. Similarly, for ScH$_{3}$, Raman and infrared\cite{PhysRevB.84.064132} studies observed an hcp-intermediate-cubic phase at $25$~GPa. Theoretically, Kong \textit{et al.}\cite{Kong_2013} reported the hcp-cubic phase transition at $25$~GPa, while Pakornchote \textit{et al.}\cite{Pakornchote_2013} found the cubic phase to be energetically more stable at slightly lower pressure of $22$~GPa. For these metal-hydrides, YH$_{3}$ and ScH$_{3}$, Kim \textit{et al.}\cite{Kim2793,BI2019} performed first principles calculations and found that the fcc structure is dynamically stable on both of them at a pressure of $\approx 18$~GPa, with an estimated $T_{c}$ around $18$ and $40$~K, respectively, with a rapid decreased of $T_{c}$ as the pressure is incremented, in agreement with the experimental report of Kong \textit{et al.}\cite{YH6em} where superconductivity where not found above $5$~K for metallic fcc structure.

Besides applied pressure, doping is another mechanism for metallization of metal hydrides, in order to induce or increase superconductivity by the improvement of some properties, like the electronic density of states at the Fermi level ($N(0)$) or the el-ph coupling. For example, the substitution of Li by Be, Mg, or Ca in LiH was studied by Zhang \textit{et al.}\cite{Zhang_2007}. In that work, the dopant acts as a donor which delivers electrons to the system, obtaining a $n$-doped material with a $T_{c}=7.78$~K for an electron content as high as 2.06, calculated at ambient pressure. Another work on that direction was performed by Olea-Amezcua \textit{et al.}\cite{PhysRevB.99.214504}. There the authors showed the metallization of alkali-metal hydrides LiH, NaH, and KH by doping with alkaline-earth metals Be, Mg, and Ca, respectively, and analyzed the superconducting properties as a function of metal content. The maximum estimated $T_{c}$ values were $2.1$~K for Li$_{0.95}$Be$_{0.05}$H, $28$~K for Na$_{0.8}$Mg$_{0.2}$H, and even $49$~K for K$_{0.55}$Ca$_{0.45}$H, without applied pressure. More recently, Villa-Cort{\'e}s \textit{et al.} \cite{Villa_Cort_s_2021} studied the structural, electronic, and lattice dynamical properties, as well as the electron-phonon coupling and superconducting critical temperature ($T_c$) of ScH$_2$ and YH$_2$ metal hydrides solid solutions, as a function of the electron- and hole-doping content, in absence of applied pressure. They found that for electron-doping content $x > 0.5$, $T_c$ increases rapidly, reaching its maximum value of the entire range at the Sc$_{0.05}$Ti$_{0.95}$H$_2$ and Y$_{0.2}$Zr$_{0.8}$H$_2$ solid solutions. These results show that such scheme to induce metallization and superconductivity on metal-hydrides works as an alternative to the applied-pressure approach. So, in this paper we implement it, in addition to applied pressure, in order to study the ScH$_{3}$ and YH$_{3}$ compounds in the fcc structure (NaCl (B1) phase), which is reported to present superconductivity, as discussed previously. Under this approach we are able to trace down the evolution of the structural, electronic, and lattice dynamical properties, as well as the el-ph coupling and $T_c$, as a function of metal content, by inducing holes ($p$-doped) and electrons ($n$-doped), as well as applied pressure, of the proposed systems. Such approach is done by the construction of solid solutions with the metal atom of the hydride: Sc$_{1-x}$M$_{x}$H$_{3}$ (M=Ca,Ti) and Y$_{1-x}$M$_{x}$H$_{3}$ ($M$=Sr,Zr)  within the Density Functional Theory (DFT)\cite{PhysRev.140.A1133}, using the virtual crystal approximation (VCA)\cite{vca}, which has been successfully applied on the study of doped superconductors \cite{PhysRevLett.88.127001,PhysRevLett.93.237002,PhysRevB.93.224513,PhysRevB.79.134523,PhysRevB.99.214504}. 

The paper is organized as follows. The computational details of our method are presented in Section II. In Section III.A we present our results related to the structural properties; while in Section III.B the electronic structure analysis is shown. The lattice dynamics are discussed in Section III.C; and the electron-phonon and superconducting properties, as well as $T_{c}$, are shown in Section III.D. Last, our conclusions are presented in Section IV.

\section{Computational details}

The calculations of the structural, electronic, lattice dynamics and el-ph coupling properties were carried out within the framework of Density Functional Theory (DFT)\citep{PhysRev.140.A1133} and Density Functional Perturbation Theory (DFPT) \citep{0953-8984-21-39-395502,RevModPhys.73.515}, both implemented in the QUANTUM ESPRESSO suit code \citep{0953-8984-21-39-395502}. The calculations were performed with a $24\times24\times24$ $k$-point mesh, and a $60$~Ry cutoff for the plane-wave basis, while the Perdew-Burke-Ernzerhof (PBE) functional \citep{PhysRevLett.77.3865} was employed to take into account the exchange and correlation contributions. 

Fourier interpolation of dynamical matrices, calculated on a $8\times8\times8$ $q$-point mesh, were used to determine the complete phonon spectra of the studied systems. Corrections due to quantum fluctuations at zero temperature, zero-point energy (ZPE) effects, are estimated through the quasi-harmonic approximation (QHA)\citep{10.2138/rmg.2010.71.3,PhysRevB.99.214504}. Within this approximation, the phonon contribution to the ground-state energy is taken into account and then a equation of state for each concentration $x$ can be constructed. Thus, the electronic structure, lattice dynamics, and el-ph properties, calculated for the fcc (B1) crystal structure at different applied pressure values, incorporate the ZPE correction.

\begin{figure}
\includegraphics[width=8.4cm]{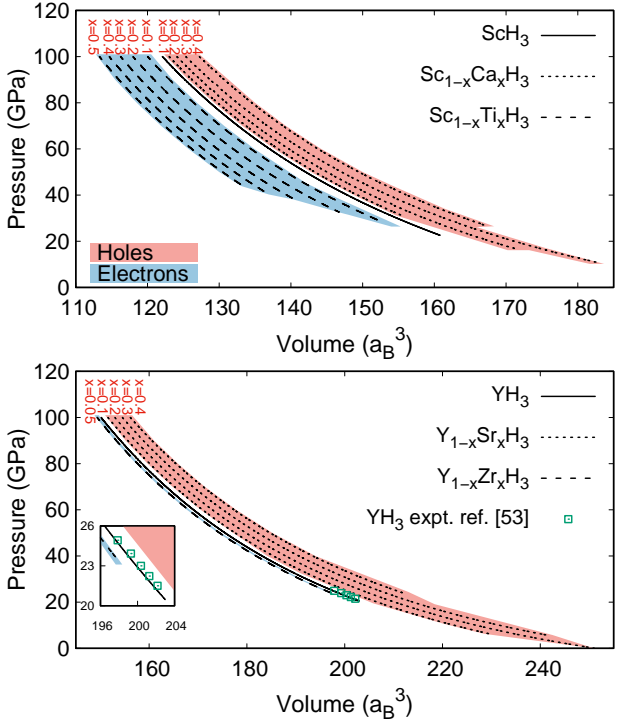}
\caption{\label{fig:pre_vol} $PV$ equation of state of Sc$_{1-x}$M$_{x}$H$_{3}$ and Y$_{1-x}$M$_{x}$H$_{3}$, for different metal M content ($x$), studied within the ZPE scheme. $a_B$ stands for the Bohr radius.}
\end{figure}

Included on the el-ph calculations, the phonon linewidths of the $\vec{q}\nu$ phonon mode $\gamma_{\vec{q}\nu}$ were also obtained, which are given by\citep{PhysRevB.6.2577,PhysRevB.9.4733}

\begin{equation}
\gamma_{\vec{q}\nu}=2\pi\omega_{\vec{q}\nu}\sum_{\vec{k}nm}\left|g_{\vec{k}+\vec{q},\vec{k}}^{\vec{q}\nu,nm}\right|^{2}\delta\left(\epsilon_{\vec{k}+\vec{q},m}-\epsilon_{F}\right)\delta\left(\epsilon_{\vec{k},n}-\epsilon_{F}\right),
\end{equation}
where $g_{\vec{k}+\vec{q},\vec{k}}^{\vec{q}\nu,nm}$ are the matrix elements of the electron-phonon interaction (calculated over a denser $48\times48\times48$ $k$-point mesh), $\epsilon_{\vec{k}+\vec{q},m}$ and $\epsilon_{\vec{k},n}$ are one-electron band energies, with band index $m,n$, and vectors $\vec{k}+\vec{q},~\vec{k}$, respectively, while $\omega_{\vec{q}\nu}$ is the phonon frequency for mode $\nu$ at wave-vector $\vec{q}$.

\begin{figure}
\includegraphics[width=8.4cm]{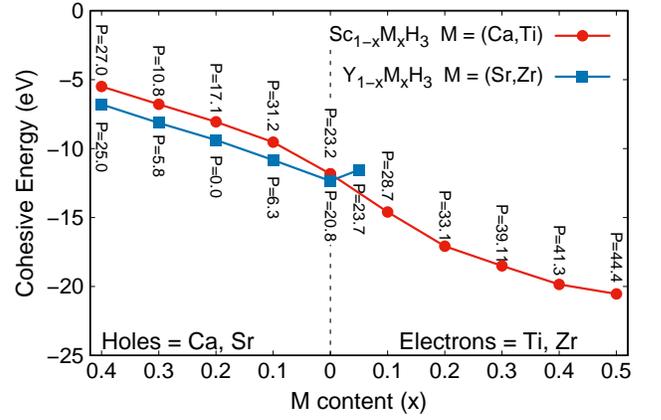}
\caption{\label{fig:cohe}Cohesive energy of Sc$_{1-x}$M$_{x}$H$_{3}$ and Y$_{1-x}$M$_{x}$H$_{3}$ as a function of the metal M content ($x$) for the minimum pressure value in the stability range of each $x$.}
\end{figure}

With the knowledge of $\gamma_{\vec{q}\nu}$ and $\omega_{\vec{q}\nu}$, the isotropic Eliashberg spectral function, $\alpha^{2}F\left(\omega\right)$, can be described as 

\begin{equation}
\alpha^{2}F\left(\omega\right)=\frac{1}{2\pi\hbar N\left(0\right)}\sum_{\vec{q}\nu}\delta\left(\omega-\omega_{\vec{q}\nu}\right)\frac{\gamma_{\vec{q}\nu}}{\omega_{\vec{q}\nu}},
\end{equation}
where $N\left(0\right)$ is the electronic density of states (per atom and spin) at $\epsilon_{F}$, and a sum over a denser Fourier interpolated $q$-point mesh of $54\times54\times54$ was required. In addition, the average electron-phonon coupling constant $\lambda$, which quantifies the coupling strength as well as the Allen-Dynes characteristic phonon frequency $\omega_{ln}$\citep{PhysRevB.12.905}, are related to the Eliashberg function as 

\begin{equation}
\lambda=2\int_{0}^{\infty}d\omega\frac{\alpha^{2}F\left(\omega\right)}{\omega}=\frac{1}{2\pi\hbar N\left(0\right)}\sum_{\vec{q}\nu}\frac{\gamma_{\vec{q}\nu}}{\omega_{\vec{q}\nu}^{2}},
\label{eq:lambda}
\end{equation}

\begin{equation}
\omega_{ln}=\exp\left\{ \frac{2}{\lambda}\int_{0}^{\infty}d\omega\frac{\alpha^{2}F\left(\omega\right)}{\omega}\ln\omega\right\}.
\end{equation}

Regarding $T_{c}$, it was estimated for each case by solving numerically the isotropic Migdal-Eliashberg gap equations\citep{Eliashberg,Bergmann1973,VILLACORTES2018371}, using the respective $\alpha^{2}F\left(\omega\right)$ for each content $x$ at its specific applied pressure value, and treating the Coulomb pseudopotential as an adjusted parameter.

Furthermore, we used the formalism of Rainer and Culleto for the calculation of the differential isotope effect coefficient \citep{PhysRevB.19.2540,VILLACORTES2022110451}, $\beta\left(\omega\right)$, to gain more insight into the coupling and how a specific phonon-frequency interval contributes to $T_{c}$. $\beta\left(\omega\right)$ is defined as 

\begin{equation}
\beta\left(\omega\right)\equiv R\left(\omega\right)\alpha^{2}F\left(\omega\right),
\label{eq:dif_alf}
\end{equation}

where $R\left(\omega\right)$ is given by 

\begin{figure*}
\includegraphics[width=18cm]{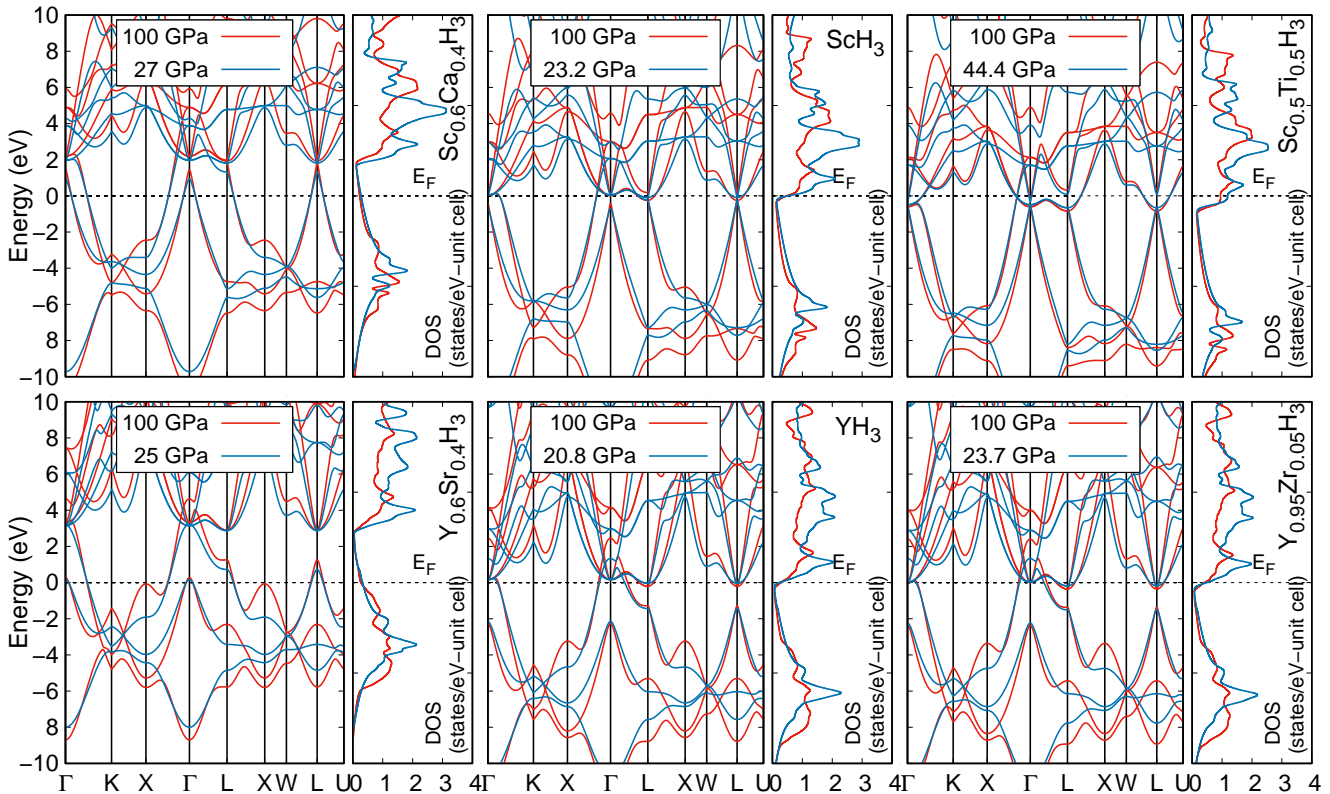}
\caption{\label{fig:bandas}Electronic band structure and density of states (DOS), for Sc$_{1-x}$M$_{x}$H$_{3}$ and Y$_{1-x}$M$_{x}$H$_{3}$ for the pristine ($x=0$), the threshold electron- (M=Ti,Zr) and hole- (M=Ca,Sr) content, each of them at the minimum pressure where the system is dynamically stable and the maximum pressure considered.}
\end{figure*}

\begin{equation}
R\left(\omega\right)=\frac{d}{d\omega}\left[\frac{\omega}{2T_{c}}\frac{\delta T_{c}}{\delta\alpha^{2}F\left(\omega\right)}\right],\label{eq:we}
\end{equation}
and the $T_{c}$ functional derivative respect to the Eliashberg function is expressed as \citep{Bergmann1973,VILLACORTES2018371,meromio,VILLACORTES2022110451} 

\begin{equation}
\frac{\delta T_{c}}{\delta\alpha^{2}F\left(\omega\right)}=-\left.\left(\frac{d\rho}{dT}\right)\right|_{T_{c}}\frac{\delta\rho}{\delta\alpha^{2}F\left(\omega\right)}, \label{eq:dev_a2f}
\end{equation}
with $\rho$ corresponding to the breaking parameter that becomes zero at $T_{c}$. Finally, the isotope effect coefficient, $\alpha$, in a specific frequency interval is given by 

\begin{equation}
\alpha=\int_{\omega_{1}}^{\omega_{2}}d\omega\beta\left(\omega\right).\label{eq:IntAlfa}
\label{eq:alf}
\end{equation}

\section{Results and discussion}

\subsection{Structural properties}

We performed structural optimizations of the fcc (cubic NaCl (B1)) structure ($Fm\bar{3}m$ space group) with a primitive cell of four atoms (one metal and three hydrogen) for the two Sc$_{1-x}$M$_{x}$H$_{3}$ (M=Ca, Ti) and Y$_{1-x}$M$_{x}$H$_{3}$ (M=Sr, Zr) solid solutions at different values of metal M content $x$. For both systems, the equations of state were constructed from the minimum pressure, at which each system becomes stable at the cubic structure, to 100 GPa as maximum pressure value.

For Sc$_{1-x}$M$_{x}$H$_{3}$, the equation of state was determined for concentrations up to $x=0.5(0.4)$ of electron(hole) doping, while for Y$_{1-x}$M$_{x}$H$_{3}$, the range was for concentrations up to $x=0.05(0.4)$ of electron(hole) doping. Electron- and hole-doping thresholds and the minimum pressures where the systems are stable were determined through dynamical instabilities, observed as imaginary frequencies in the phonon dispersion for larger $x$ contents and smaller pressure values. Phonon instabilities in metal hydrides induced by alloying have been reported previously in literature \citep{doi:10.1063/1.4714549,PhysRevB.69.094205,PhysRevB.99.214504}, where such dynamical behavior have been related to an increase of the heat of formation, meaning that the solid solutions become less stable.

In Fig. \ref{fig:pre_vol}, we show the evolution of the equation of state of each metal content $x$. For both systems, Sc$_{1-x}$M$_{x}$H$_{3}$ and Y$_{1-x}$M$_{x}$H$_{3}$, increasing the electron content leads to a monotonous reduction of the volume, at a given pressure, as well as an increment in the minimum stable pressure value. For hole doping, the volume tendency is opposite, while the minimum stable pressure values are not following any specific trend. This behavior indicates a strengthening of the chemical bonding as the electron-content is increased, given by the increment of Zr- and Ti-content, suggesting a hardening of phonon frequencies. Similarly, as the hole-content grows, by the increase of Sr- and Ca-content, the chemical bonding gets weaker, implying a softening of the phonon frequencies. A complete set of structural parameters, that is, the equilibrium volume ($V_{0}$), bulk modulus ($B_{0}$), and its pressure derivative ($B^{'}_{0}$) are given in Tab. \ref{tab:table1} (Appendix \ref{sec:stateequation}), for both systems, of all M content ($x$) values studied. It is worth noting that our YH$_{3}$ results are in excellent agreement with the experimental data reported by Machida \textit{et al.}\cite{MACHIDA2006436}.

In Fig.\ref{fig:cohe} we show the calculated cohesive energy ($E_{coh}$) of the two systems, within their respective electron and hole stability-range at their corresponding minimum pressure threshold. This quantity is used to characterize the stability of alloys and solid solutions, and is given by the following:

\begin{equation}
E_{coh}=E_{sys}^{tot}-(1-x)E_{N}^{a}-xE_{M}^{a}-3E_{H}^{a},
\end{equation}
where $E_{sys}^{tot}$ is the total energy of the N$_{1-x}$M$_{x}$H$_{3}$ solid solution at content $x$, while $E_{N}^{a}$, $E_{M}^{a}$, and $E_{H}^{a}$ are the calculated total energies of the isolated atoms N = Y, Sc; M = Sr, Zr, Ca, Ti; and hydrogen, respectively. In general, for the two solid solutions, the hole-doped systems are less stable than their corresponding pristine systems ($x=0$) (the larger the $E_{coh}$ absolute value, the more stable the system is), nevertheless, they are still in the stability range (negative $E_{coh}$). For the case of electron-doped regime, while Y$_{1-x}$Zr$_{x}$H$_{3}$ follows the same observed tendency than the hole-doped systems, we found that Sc$_{1-x}$Ti$_{x}$H$_{3}$ is more stable than the pristine one, indicating the possibility to synthesize experimentally such solid solutions.

With the optimized lattice parameters and the corresponding equation of state for each system, at different content for their electron- and hole-doping regions, we proceeded to calculate their electronic and lattice dynamical properties for different applied pressure values at each $x$. Furthermore, we are presenting results obtained by the ZPE scheme. While the ZPE effects on the electronic properties are hardly visible, comparing with the static scheme, on the lattice dynamical ones, there is a noticeable softening as a general effect. This tendency comes mainly from the unit cell expansion as the ZPE contribution to the energy is taken into account.

\begin{figure}
\includegraphics[width=8.4cm]{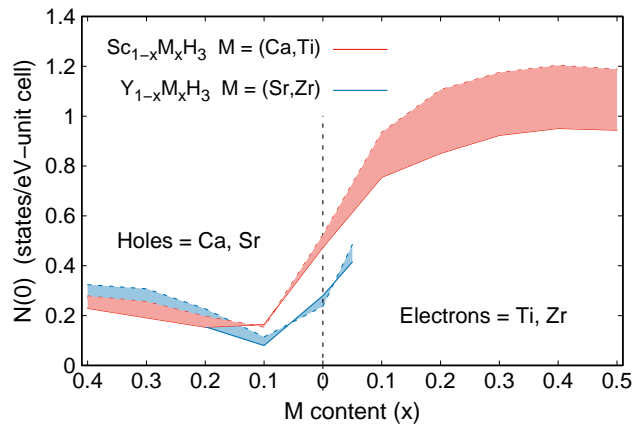}\caption{\label{fig:EFermi}
Evolution of the total density of states at the Fermi level, $N\left(0\right)$, for Sc$_{1-x}$M$_{x}$H$_{3}$ and Y$_{1-x}$M$_{x}$H$_{3}$ as a function of the M content $x$ spanning the range of studied applied pressure.}
\end{figure}

\subsection{Electronic properties}

\begin{figure*}
\includegraphics[width=18cm]{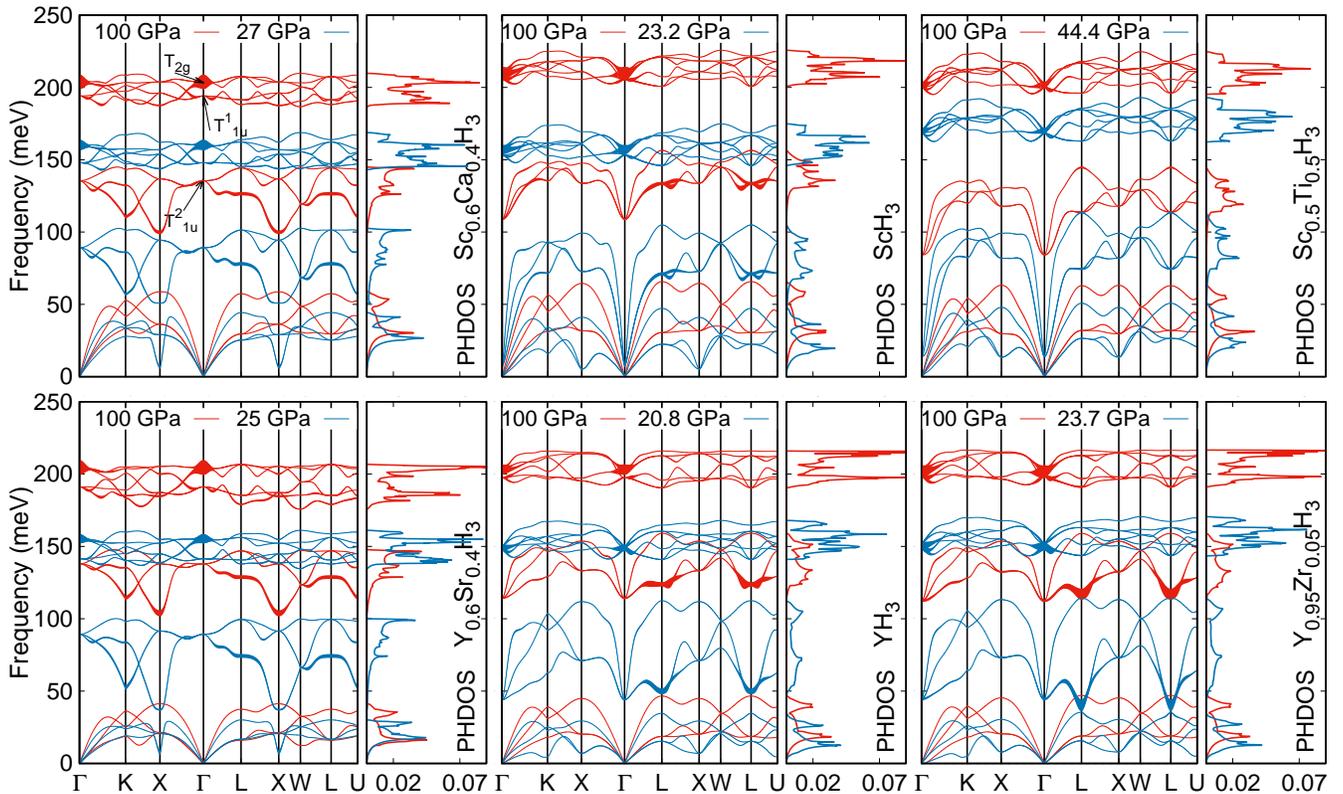}
\caption{\label{fig:phonons}  Phonon dispersion, linewidths (as vertical lines along the phonon branches) and PHDOS for Sc$_{1-x}$M$_{x}$H$_{3}$ and Y$_{1-x}$M$_{x}$H$_{3}$ at the pristine and threshold electron- (Ti,Zr) and hole- (Ca,Sr) content, each of them at their corresponding limit pressure values.}
\end{figure*}

The evolution of the electronic band structure and the density of states at the Fermi level, $N\left(0\right)$, is analyzed in order to evaluate the effects of electron- and hole-content and pressure on the electronic properties of the solid-solutions.

In Fig. \ref{fig:bandas} we show the band structure for Sc$_{1-x}$M$_{x}$H$_{3}$ and Y$_{1-x}$M$_{x}$H$_{3}$ of the pristine system and of their corresponding threshold electron- (Ti, Zr) and hole- (Ca, Sr) doping content, each of them at the minimum pressure (where the system is dynamically stable) and the maximum pressure considered ($100$~GPa). 

It can be seen that the main pressure effect on the band structure, independent of the electron- or hole-content, is a slight increase of the bandwidth. That is, the low-energy bands are push to even lower energy, and the high-energy ones to higher values, while keeping almost unaffected the bands around the Fermi level ($E_F$). The band structure of the pristine ScH$_3$ shows a threefold degenerate state on the L-point (giving place to one hole-like and two electron-like bands) and a twofold degenerate state on the $\Gamma$-point (electron-like bands), both located at $E_F$. In comparison, the pristine YH$_3$ compound shows a twofold state on the L-point, also at the Fermi level, while the twofold state on the $\Gamma$-point is located just above $E_F$. As the electron content is increased for ScH$_3$ (by Ti), the degeneracy at the L-point breaks, giving place to a new twofold state that continue to move far away from $E_{F}$, rising as an electron-like band and a hole-like band. Instead, at $\Gamma$-point the twofold state is intact, moving as a whole to lower energy values, indicating that electronic topological transitions (ETT) take place. In the case of electron-doping for YH$_3$ (by Zr), the degenerate states at L- and $\Gamma$-points are maintained, while they undergo a minor shift to lower energies. Regarding the hole-content increase on both systems, SH$_3$ by Ca and YH$_3$ by Sr, its essential effect is the shift of the band structure to higher energies, without noticeable changes on the degenerate states discussed before.

Analyzing the evolution of $N\left(0\right)$, as a function of M content ($x$) for the minimum and maximum pressure values of both systems, Sc$_{1-x}$M$_{x}$H$_{3}$ and Y$_{1-x}$M$_{x}$H$_{3}$ (see Fig. \ref{fig:EFermi}), it can be observed that, in general, $N(0)$ reduces at $x=0.1$ on the hole-doping regime, and in a more drastic way for the ScH$_3$ case. As the hole-content increases, $N(0)$ starts to rise slowly, getting close to its value at $x=0$ for the YH$_3$ system. For the electron doping regime, while YH$_3$ undergoes a minor increase, mainly by the threshold limit on the Zr-content, $N(0)$ of ScH$_3$ grows rapidly, doubling its pristine value at $x(\mbox{Ti})=0.3$. As $x$ increases even more, the $N(0)$ grow ratio slows considerably, tending to saturate. As the applied pressure raises, $N(0)$ shows an small reduction, which is due to the expand of the bandwidth discussed previously.

\subsection{Lattice dynamics} 

The phonon dispersion is presented on Fig. \ref{fig:phonons}, including their respective phonon linewidth $\gamma_{\vec{q}\nu}$ and the phonon density of states (PHDOS), for Sc$_{1-x}$M$_{x}$H$_{3}$ and Y$_{1-x}$M$_{x}$H$_{3}$ at the pristine ($x=0$) and the threshold electron- (Ti,Zr) and hole- (Ca,Sr) contents. In general, for both systems, the optical branches soften slightly as the hole-content increases, while they are shifted to higher frequencies as the electron-content rises. In particular for the high-frequency optical branches, while they shift almost on a rigid way above the frequencies of the pristine systems for the electron-doping case, on the hole-doping solid-solutions they show, in addition to the softening, a small renormalization in the $T_{2g}$ and $T^{1}_{1u}$ optical branches at $\Gamma$. While the mid-frequency region shows subtle renormalizations at $\Gamma$- and L-point for electron-content regime, they are stronger for the same high-symmetry points, in addition to the X-point, for hole-content systems. Regarding the acoustic branches, they remain almost unchanged for electron-doping, whereas for hole-doping they lightly harden.  
Interestingly, the phonon linewidths $\gamma_{\vec{q}\nu}$ (vertical lines along the phonon branches), mainly localized around $\Gamma$ at the $T_{2g}$ and $T^{1}_{1u}$ optical phonon branches for $x=0$, remain almost unchanged for the hole-doping regime in both alloys, while for the electron-doping regime it slightly reduces in Sc$_{1-x}$M$_{x}$H$_{3}$, and increases in Y$_{1-x}$M$_{x}$H$_{3}$. In general, for both solid solid-solutions, independent of doping-scheme, the observed effect of applied pressure is a rigid hardening of the phonon frequencies and a lifting of phonon anomalies, that leans to instabilities at K and X-point, for the acoustic branches.

\subsection{Electron-phonon and superconducting properties} 

\begin{figure}
\includegraphics[width=8.4cm]{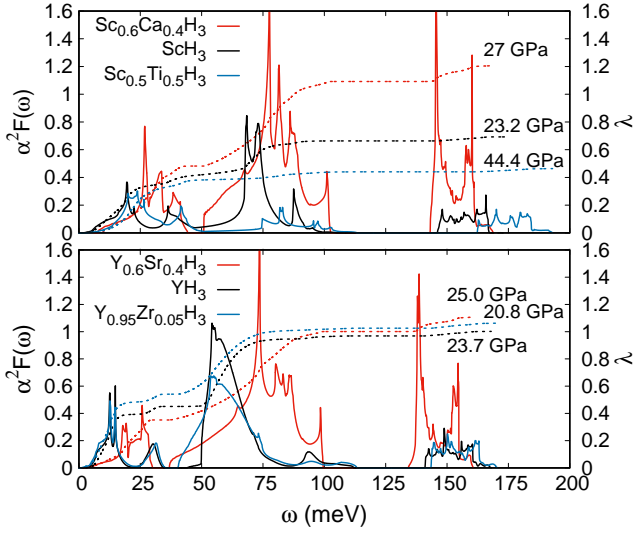}
\caption{\label{fig:A2f}Eliashberg function and partial integrated electron-phonon coupling parameter $\lambda(\omega)$ for Sc$_{1-x}$M$_{x}$H$_{3}$ and Y$_{1-x}$M$_{x}$H$_{3}$ at $x=0$ and at the threshold electron- and hole-content for each solid-solution at the minimum pressure where the systems are dynamically stable.}
\end{figure}

With the electronic and the lattice dynamics information, the electron-phonon spectral functions $\alpha^{2}F\left(\omega\right)$ were calculated for the entire range of hole- and electron-content stable regimes in a broad pressure range. The $\alpha^{2}F\left(\omega\right)$ for the threshold electron- and hole-content, as well as the pristine cases, at the minimum pressure where the systems are dynamically stable, can be seen in Fig. \ref{fig:A2f}. As the electron(hole)-content increases on both systems, the high-frequency optical region of the Eliashberg function shifts to higher(lower) frequencies, while the mid-range frequency region shows an opposite behavior, and the acoustical one practically does not shift. For both systems, the weight of the spectral function is incremented as the doping goes from the electron- to the hole-content thresholds, passing through the pristine one ($x=0$). Regarding the pressure effects, in general, for all different $x$ content cases (electron- and hole-doping), at both solid-solutions, the spectra shift to a higher frequency region as the pressure arises, going from the minimum dynamically stable value up to 100 GPa. 

As the Eliashberg spectral function determines the electron-phonon coupling parameter $\lambda$ (see Eq. \ref{eq:lambda}), the $\alpha^2F(\omega)$ observed shift due to both, doping and applied pressure, has an impact on $\lambda$ as well. The evolution of the electron-phonon coupling constant as a function of frequency, $\lambda(\omega)$, is shown in Fig.\ref{fig:A2f}. For the pristine ($x=0$) and the electron-doped Sc$_{1-x}$Ti$_{x}$H$_{3}$ solid solution, it can be observed that the main contribution to $\lambda$ comes from the acoustic region. However, for the hole-doped Sc$_{1-x}$Ca$_{x}$H$_{3}$, the main contribution comes from the mid-range frequency optical phonons. The behavior of $\lambda(\omega)$ for the Y$_{1-x}$M$_{x}$H$_{3}$ solid solution is slightly different. In this case, both regions, the acoustic and the mid-range frequency optical ones, contribute almost at the same rate to $\lambda(\omega)$. It is worth noting that, for both solid-solutions, the high-frequency optical phonons have marginal contribution to $\lambda$. 

\begin{figure}
\includegraphics[width=8.4cm]{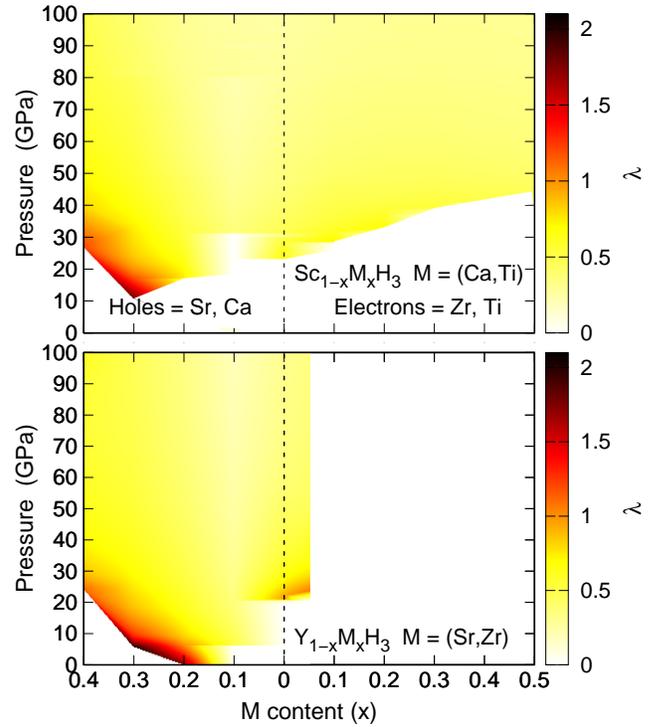}
\caption{\label{fig:lamb}Electron-phonon coupling constant ($\lambda$) of Sc$_{1-x}$M$_{x}$H$_{3}$ and Y$_{1-x}$M$_{x}$H$_{3}$ for the entire studied range of electron- and hole-content and applied pressure for each solid-solution. }
\end{figure}

In order to analyze the behavior of $\lambda$ as a function of pressure and electron- or hole-content for the solid-solutions, we present it as a color-map plot in Fig. \ref{fig:lamb}. In general, it can be observed that for most of the scanned pressure values, on both doping schemes (electron and hole) $\lambda$ hardly goes beyond $1$, regardless of the solid-solution. It is worth to mention that higher $\lambda$ values are obtained for pressure close to dynamical instabilities at each electron- or hole-threshold content. Interestingly, only for very specific combination of pressure and hole-content it is possible to reach higher $\lambda$ values, like $1.8$ for Sc$_{0.7}$Ca$_{0.3}$H$_{3}$ at $10.8$~GPa, and $2.0$ for Y$_{0.7}$Sr$_{0.3}$H$_{3}$ at $5.8$~GPa. In particular for the latter solid-solution, the region of pressure and hole-content that could provide $\lambda$ values close to $2$ is spread between $x=0.2$ and $0.3$, and around $1$ to $6$~GPa on applied pressure.

\begin{figure}
\includegraphics[width=8.4cm]{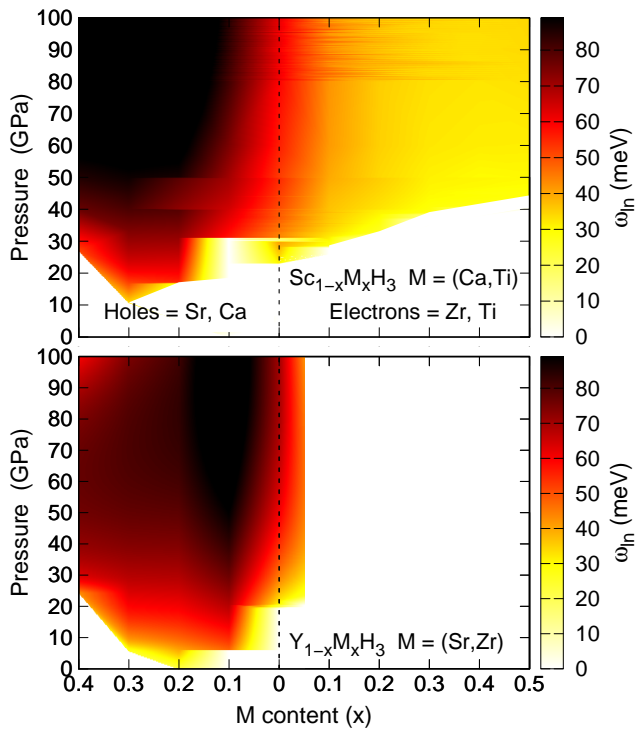}
\caption{\label{fig:wln} Allen-Dynes characteristic phonon frequency ($\omega_{ln}$) for Sc$_{1-x}$M$_{x}$H$_{3}$ and Y$_{1-x}$M$_{x}$H$_{3}$ of the complete range of electron- and hole-content and applied pressure for each solid solution.}
\end{figure}

In a similar fashion as $\lambda$, the evolution of the Allen-Dynes characteristic phonon frequency, 
$\omega_{ln}$, as a function of pressure and electron- and hole-content is shown in Fig. \ref{fig:wln}. It can be observed a hardening as the pressure increases, specially noticeable for the hole-doping regime and more subtle for the electron-doping one, while the lower $\omega_{ln}$ values are located at the pressure and hole-content regions where both solid-solutions present their maximum $\lambda$ values.

\begin{figure}
\includegraphics[width=8.4cm]{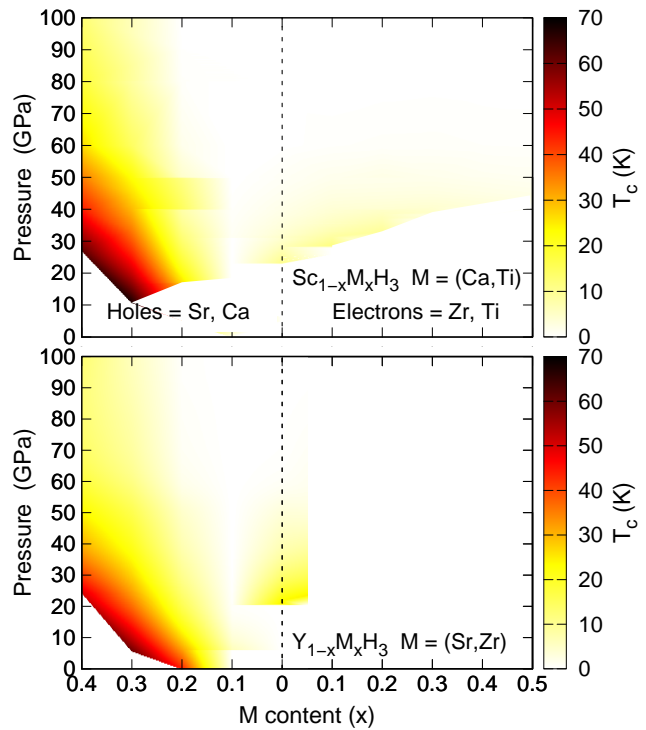}
\caption{\label{fig:TCP} Superconducting critical temperature, $T_{c}$, calculated with $\mu^*=0.15$, of Sc$_{1-x}$M$_{x}$H$_{3}$ and Y$_{1-x}$M$_{x}$H$_{3}$ at the entire range of electron- and hole-content and applied pressure for each solid-solution.}
\end{figure}

The calculated electron-phonon coupling properties were used to obtain estimates for the superconducting critical temperature, $T_c$, as a function of applied pressure and content $x$ for both solid solutions. The isotropic Migdal-Eliashberg gap equations where solved numerically with two different values of the Coulomb pseudopotential ($\mu^{*}$): $\mu^{*}=0$ (which provides an upper limit for $T_{c}$) and $0.15$, in order to get an idea of how strong $T_c$ could be affected by the $\mu^{*}$ variation. In general, we get the maximum $T_{c}$ at the minimum pressure values were we found dynamically stable solid solutions for each content ($x$) of both regimes, electron- and hole-doping, as can be observed in Fig. \ref{fig:TCP}. Comparing doping regimes, the hole-doping one reports the highest critical temperature, with values of $T_{c}=92.7(67.9)$~K and $T_{c}=84.5(60.2)$~K at $x=0.3$ for Sc$_{1-x}$Ca$_{x}$H$_{3}$ and Y$_{1-x}$Sr$_{x}$H$_{3}$ respectively, with $\mu^{*}=0(0.15)$. These maximum $T_c$ values corresponds to the highest $\lambda$, the lowest $\omega_{ln}$, and a comparatively low $N(0)$ (related to its corresponding electron-doping values). Such behavior shows that the tuning of the lattice dynamics is the path to enhance superconductivity in both systems. It is worth to mention that Y$_{0.8}$Sr$_{0.2}$H$_{3}$, the only system that is dynamically stable at ambient pressure ($0$~GPa), presents $T_c=65.4(44.3)$~K for $\mu^*=0(0.15)$. 

\begin{figure}
\includegraphics[width=8.4cm]{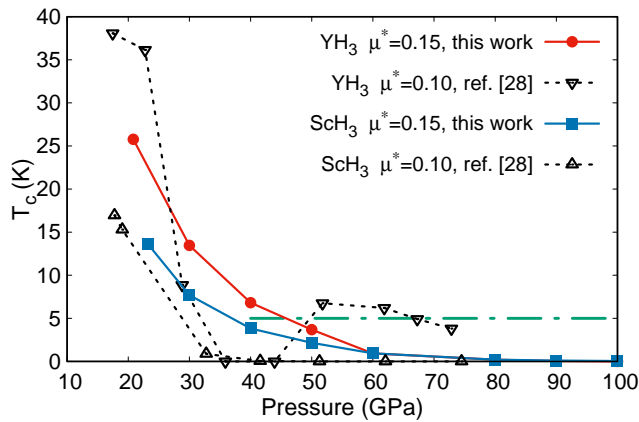}
\caption{\label{fig:compa} Superconducting critical temperature ($T_{c}$) calculated in this work at the entire range of applied pressure for YH$_3$ and ScH$_3$, compared with the calculated ones by Kim \textit{et al.}\cite{Kim2793}. The horizontal green-line represents the experimental minimal threshold temperature ($5$~K) at which Kong \textit{et al.}\cite{YH6em} searched for superconductivity for pure-fcc metallic YH$_3$, starting at $40$~GPa.}
\end{figure}

Regarding the pristine systems, we are fully aware that have not been experimental observation of superconductivity for fcc-phase YH$_3$ and SH$_3$ compounds. In particular, Kong \textit{et al.}\cite{YH6em} reported no superconductivity for temperatures above $5$~K for pure-fcc metallic YH$_3$, at pressures values going from $40$~GPa up to $180$~GPa. Then, in Fig. \ref{fig:compa} we show our $T_c$ results for ScH$_{3}$ and YH$_{3}$, as a function of applied pressure, and from there it can be seen that for pressure values above $40$~GPa, $T_c$ goes below $5$~K, and gets even smaller as pressure arises, in good agreement with the experimental reports. Additionally, comparing with previous results calculated by Kim \textit{et al.}\cite{Kim2793}, it can be observed that the downward tendency is reproduced on both systems.

Finally, to get more insight about the contribution of the different phonon-frequency intervals to the superconductivity state, we have calculated the Rainer and Culleto differential isotope effect coefficient, $\beta\left(\omega\right)$ (Eq. \ref{eq:dif_alf}), and the partial isotope effect $\alpha(\omega)$ (Eq. \ref{eq:alf}), which give information on how strong can $T_{c}$ be modified due to an infinitesimal ion-mass change in a specific phonon interval. 

\begin{figure}
\includegraphics[width=8.4cm]{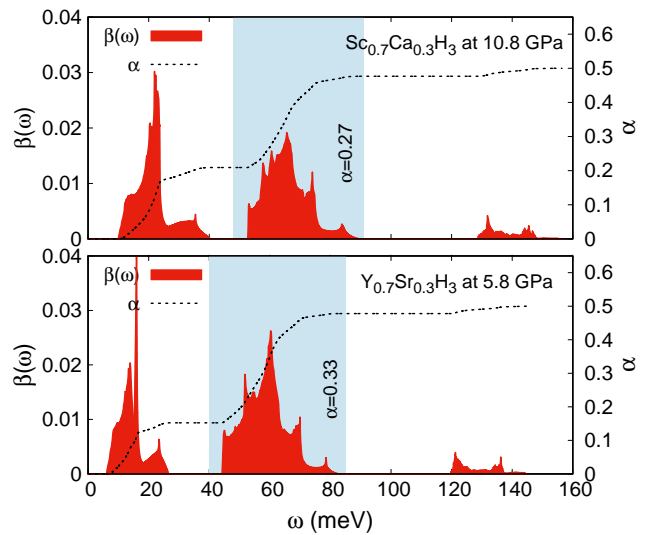}
\caption{\label{fig:RC} Differential isotope effect coefficient $\beta\left(\omega\right)$, and partial isotope effect $\alpha(\omega)$ for both solid-solutions at the pressure and hole-content that presents the highest $T_{c}$, calculated with $\mu^*=0.15$.}
\end{figure}

In Fig. \ref{fig:RC} we present $\beta(\omega)$ and $\alpha(\omega)$ for both solid-solutions at the pressure and hole-content that presents the highest $T_{c}$ ($\mu^{*}=0.15$): Sc$_{0.7}$Ca$_{0.3}$H$_{3}$ at $10.8$~GPa and Y$_{0.7}$Sr$_{0.3}$H$_{3}$ at $5.8$~GPa. We found $\alpha(\omega)$ to be $0.21(0.15)$  at the acoustic interval, and $0.27(0.33)$ for the mid-frequency region (shadow interval) for Sc$_{0.7}$Ca$_{0.3}$H$_{3}$(Y$_{0.7}$Sr$_{0.3}$H$_{3}$), values that contribute to the $41(30)\%$ and $53(65)\%$, respectively, of the total isotope coefficient $\alpha=0.49$, showing the importance of the mid-frequency region to the superconducting state. 

In order to get an idea of how these different phonon regions contribute to $T_{c}$, we calculated the critical-temperature change, $\Delta T_{c}$, when a phonon region is suppressed by means of the Bergmann and Rainer formalism\cite{Bergmann1973,VILLACORTES2018371}:

\begin{equation}
\Delta T_{c}=\int_{0}^{\infty}d\omega\frac{\delta T_{c}}{\delta\alpha^{2}F\left(\omega\right)}\Delta\alpha^{2}F\left(\omega\right),
\label{eq:tc-el-ph}
\end{equation}

where $\Delta\alpha^{2}F\left(\omega\right)=\alpha^{2}F^{'}\left(\omega\right)-\alpha^{2}F\left(\omega\right)$. Here, $\alpha^{2}F\left(\omega\right)$ is the total Eliashberg function and $\alpha^{2}F^{'}\left(\omega\right)$ is the one with a specific phonon region suppressed. Then, applying it to the same cases as before (the ones with the highest obtained $T_c$ for $\mu^{*}=0.15$) we found $\Delta T_{c}$ to be $-18.7(-11.1)$~K for the acoustic region and $-52.3(-56.6)$~K for the mid-frequency region for Sc$_{0.7}$Ca$_{0.3}$H$_{3}$(Y$_{0.7}$Sr$_{0.3}$H$_{3}$), representing the later a reduction in $T_{c}$ of $77(94)\%$ when the mid-frequency region is suppressed in $\alpha^{2}F\left(\omega\right)$.

\section{CONCLUSIONS}
We have performed a thorough analysis of the structural, electronic, lattice dynamics, electron-phonon coupling, and superconducting properties of the metal-hydride fcc solid-solutions 
Sc$_{1-x}$M$_{x}$H$_{3}$ (M=Ca,Ti) and Y$_{1-x}$M$_{x}$H$_{3}$ (M=Sr,Zr), as a function of electron- and hole-doping content $x$, as well as applied pressure. For both systems, while increasing the electron content leads to an increment of the minimum stable pressure in comparison to pristine systems, for hole-doping we found lower minimum stable pressure values for most of content $x$, even finding a particular hole-doping case, Y$_{0.8}$Sr$_{0.2}$H$_{3}$, that was dynamically stable at zero applied pressure.
Although $N(0)$ is not improved in the whole hole-doping region for both systems, and even decreasing for Sc$_{0.9}$Ca$_{0.1}$H$_{3}$ in comparison with ScH$_3$, at the electron-doping regime $N(0)$ shows an important increment at $x=0.3$ for the Sc-doped hydride, almost twice the corresponding value for the pristine system. As the pressure increases, $N(0)$ shows a reduction, which is related to the observed bandwidth expansion in both solid solutions. 
As for the lattice dynamics, the optical phonons soften in both systems by hole-doping, which is more pronounced for the high-frequency 
region than the the mid-frequency one, while the acoustic branches lightly harden. Meanwhile, for 
the electron-doping regime, the optical branches shift to higher values and the acoustic ones remain almost 
unchanged. In general, for both solid-solutions, independent of doping-scheme, the applied pressure effect
is a rigid hardening of the phonon frequencies and a lifting of phonon anomalies at the acoustic branches. 
Regarding the electron-phonon coupling parameter, $\lambda$, the main contribution comes from the acoustic and mid-range optical phonon branches for both solid solutions, while the high-frequency optical ones have marginal participation.
Interestingly, a remarkable improvement of $\lambda$ is observed in the $0.2-0.4$ hole-doping range and for low applied pressure values, close to dynamical instabilities, while by electron-doping, the systems do not improve such property, whatever value of applied pressure is considered. In particular, the systems can reach $\lambda$ values as high as $1.8$ for Sc$_{0.7}$Ca$_{0.3}$H$_{3}$ at $10.8$~GPa, and $2.0$ for Y$_{0.7}$Sr$_{0.3}$H$_{3}$ at $5.8$~GPa, which represent an improvement of $160\%$ and $100\%$, respectively, in comparison with the highest $\lambda$ values at their corresponding pristine systems, under applied pressure. Then, as a consequence, the maximum superconducting critical temperature, at each system, was obtained for those particular conditions of hole-doping and pressure, with values of 
$T_{c}=92.7(67.9)$~K and $T_{c}=84.5(60.2)$~K for ScH$_{3}$ and YH$_{3}$ doped systems, respectively, 
with $\mu^{*}=0(0.15)$, while a general reduction of $\lambda$ and $T_{c}$ as the applied pressure rises, independent of electron- or hole-doping content, is mainly coming from the phonon hardening. 
Finally, by calculating the Rainer and Culleto differential isotope effect coefficient, we found the mid-frequency region as the most crucial phonon zone to the superconducting state. Then, due to all the above, we can conclude that the tuning of the lattice dynamics is a promising path for improving the superconductivity on both systems.

\begin{acknowledgments}
The authors thankfully acknowledge computer resources, technical advise,
and support provided by Laboratorio Nacional de Superc{\'o}mputo del Sureste
de M{\'e}xico (LNS), a member of the CONACYT national laboratories. One
of the authors (S. Villa-Cort{\'e}s.) also acknowledges the Consejo Nacional
de Ciencia y Tecnolog{\'i}a (CONACyT, M{\'e}xico) by the support under grant
769301. 
\end{acknowledgments}

\appendix

\section{Equation of state parameters\label{sec:stateequation}}

In Table \ref{tab:table1} we show the equation of state parameters of the third-order Birch-Murnaghan equation\cite{Birch} for both solid solutions, Sc$_{1-x}$M$_{x}$H$_{3}$ and Y$_{1-x}$M$_{x}$H$_{3}$, in the fcc (NaCl (B1)) structure ($Fm\bar{3}m$ space group) within the ZPE scheme.

\begin{table}
\caption{
\label{tab:table1} 
Birch-Murnaghan fit to the equation of state. $V_{0}$ is the reference volume at zero pressure (in $a_{B}^{3}$, where $a_{B}$ is the Bohr radius), $B_{0}$ is the bulk modulus in GPa at zero pressure, and $B^{'}_{0}$ is the pressure derivative of the bulk modulus.}

\begin{ruledtabular}
\begin{tabular}{cccc}
\multicolumn{4}{c}{Sc$_{1-x}$Ca$_{x}$H$_{3}$}\tabularnewline
\hline 
 $x$ & $V_{0}$ ($a_{B}^{3}$) & $B_{0}$ (GPa) & $B^{'}_{0}$\tabularnewline
 0.1 & 191.10 & 100.72 & 3.66\tabularnewline
 0.2 & 197.03 & 92.94 & 3.70\tabularnewline
 0.3 & 202.76 & 86.40 & 3.70\tabularnewline
 0.4 & 207.74 & 82.47 & 3.74\tabularnewline
\hline
\multicolumn{4}{c}{ScH$_{3}$}\tabularnewline
\hline 
  & $V_{0}$ ($a_{B}^{3}$) & $B_{0}$ (GPa) & $B^{'}_{0}$\tabularnewline
  & 189.06 & 104.49 & 3.64\tabularnewline
\hline 
\multicolumn{4}{c}{Sc$_{1-x}$Ti$_{x}$H$_{3}$}\tabularnewline
\hline 
 $x$ & $V_{0}$ ($a_{B}^{3}$) & $B_{0}$ (GPa) & $B^{'}_{0}$\tabularnewline
 0.1 & 184.52 & 106.24 & 3.69\tabularnewline
 0.2 & 180.43 & 107.47 & 3.73\tabularnewline
 0.3 & 174.96 & 116.36 & 3.68\tabularnewline
 0.4 & 170.91 & 121.80 & 3.66\tabularnewline
 0.5 & 169.66 & 116.31 & 3.77\tabularnewline
\hline
\multicolumn{4}{c}{Y$_{1-x}$Sr$_{x}$H$_{3}$}\tabularnewline
\hline 
 $x$ & $V_{0}$ ($a_{B}^{3}$) & $B_{0}$ (GPa) & $B^{'}_{0}$\tabularnewline
 0.1 & 245.38 & 89.43 & 3.77\tabularnewline
 0.2 & 251.58 & 79.36 & 3.77\tabularnewline
 0.3 & 258.51 & 74.04 & 3.78\tabularnewline
 0.4 & 265.01 & 70.43 & 3.77\tabularnewline
\hline
\multicolumn{4}{c}{YH$_{3}$}\tabularnewline
\hline 
  & $V_{0}$ ($a_{B}^{3}$) & $B_{0}$ (GPa) & $B^{'}_{0}$\tabularnewline
  & 240.35 & 88.21 & 3.77 \tabularnewline
\hline
\multicolumn{4}{c}{Y$_{1-x}$Zr$_{x}$H$_{3}$}\tabularnewline
\hline 
 $x$ & $V_{0}$ ($a_{B}^{3}$) & $B_{0}$ (GPa) & $B^{'}_{0}$ \tabularnewline
 0.05 & 238.20 & 89.10 & 3.79 \tabularnewline
\end{tabular}
\end{ruledtabular}

\end{table}

\bibliographystyle{unsrt}
\bibliography{Art_MH3}

\begin{thebibliography}{10}

\bibitem{PhysRevLett.92.187002}
N.~W. Ashcroft.
\newblock Hydrogen dominant metallic alloys: High temperature superconductors?
\newblock {\em Phys. Rev. Lett.}, 92:187002, May 2004.

\bibitem{cor6}
Lijun Zhang, Yanchao Wang, Jian Lv, and Yanming Ma.
\newblock Materials discovery at high pressures.
\newblock {\em Nature Reviews Materials}, 2, Feb 2017.

\bibitem{Duan2019}
Defang Duan, Hongyu Yu, Hui Xie, and Tian Cui.
\newblock Ab initio approach and its impact on superconductivity.
\newblock {\em Journal of Superconductivity and Novel Magnetism}, 32(1):53--60,
  Jan 2019.

\bibitem{PhysRevB.96.100502}
K.~Tanaka, J.~S. Tse, and H.~Liu.
\newblock Electron-phonon coupling mechanisms for hydrogen-rich metals at high
  pressure.
\newblock {\em Phys. Rev. B}, 96:100502, Sep 2017.

\bibitem{BI2019}
Tiange Bi, Niloofar Zarifi, Tyson Terpstra, and Eva Zurek.
\newblock The search for superconductivity in high pressure hydrides.
\newblock In {\em Reference Module in Chemistry, Molecular Sciences and
  Chemical Engineering}. Elsevier, 2019.

\bibitem{10.1038/srep06968}
Duan Defang, Liu Yunxian, Tian Fubo, Li~Da, Huang Xiaoli, Zhao Zhonglong,
  Yu~Hongyu, Liu Bingbing, Tian Wenjing, and Cui Tian.
\newblock Pressure-induced metallization of dense (h2s)2h2 with high-tc
  superconductivity.
\newblock {\em Scientific Reports}, 4, Sept 2014.

\bibitem{Liu6990}
Hanyu Liu, Ivan~I. Naumov, Roald Hoffmann, N.~W. Ashcroft, and Russell~J.
  Hemley.
\newblock Potential high-tc superconducting lanthanum and yttrium hydrides at
  high pressure.
\newblock {\em Proceedings of the National Academy of Sciences},
  114(27):6990--6995, 2017.

\bibitem{Wang6463}
Hui Wang, John~S. Tse, Kaori Tanaka, Toshiaki Iitaka, and Yanming Ma.
\newblock Superconductive sodalite-like clathrate calcium hydride at high
  pressures.
\newblock {\em Proceedings of the National Academy of Sciences},
  109(17):6463--6466, 2012.

\bibitem{203}
A.~P. Drozdov, M.~I. Eremets, I.~A. Troyan, V.~Ksenofontov, and S.~I. Shylin.
\newblock Conventional superconductivity at 203 kelvin at high pressures in the
  sulfur hydride system.
\newblock {\em Nature, Letter}, 525:73--76, Sept 2015.

\bibitem{eina}
Mari Einaga, Masafumi Sakata, Takahiro Ishikawa, Katsuya Shimizu, Mikhail~I.
  Eremets, Alexander~P. Drozdov, Ivan~A. Troyan, Naohisa Hirao, and Yasuo
  Ohishi.
\newblock Crystal structure of the superconducting phase of sulfur hydride.
\newblock {\em Nature Physics}, 12, 2016.

\bibitem{La50}
A.~P. Drozdov, P.~P. Kong, V.~S. Minkov, S.~P. Besedin, M.~A. Kuzovnikov,
  S.~Mozaffari, L.~Balicas, F.~Balakirev, D.~Graf, V.~B. Prakapenka,
  E.~Greenberg, D.~A. Knyazev, M.~Tkacz, and M.~I. Eremets.
\newblock Superconductivity at 250 k in lanthanum hydride under high pressures.
\newblock {\em Nature}, 569:528--531, 2019.

\bibitem{PhysRevLett.122.027001}
Maddury Somayazulu, Muhtar Ahart, Ajay~K. Mishra, Zachary~M. Geballe, Maria
  Baldini, Yue Meng, Viktor~V. Struzhkin, and Russell~J. Hemley.
\newblock Evidence for superconductivity above 260 k in lanthanum superhydride
  at megabar pressures.
\newblock {\em Phys. Rev. Lett.}, 122:027001, Jan 2019.

\bibitem{PhysRevLett.126.117003}
Elliot Snider, Nathan Dasenbrock-Gammon, Raymond McBride, Xiaoyu Wang, Noah
  Meyers, Keith~V. Lawler, Eva Zurek, Ashkan Salamat, and Ranga~P. Dias.
\newblock Synthesis of yttrium superhydride superconductor with a transition
  temperature up to 262 k by catalytic hydrogenation at high pressures.
\newblock {\em Phys. Rev. Lett.}, 126:117003, Mar 2021.

\bibitem{YH6em}
Panpan Kong, Vasily~S. Minkov, Mikhail~A. Kuzovnikov, Alexander~P. Drozdov,
  Stanislav~P. Besedin, Shirin Mozaffari, Luis Balicas, Fedor~Fedorovich
  Balakirev, Vitali~B. Prakapenka, Stella Chariton, Dmitry~A. Knyazev, Eran
  Greenberg, and Mikhail~I. Eremets.
\newblock Superconductivity up to 243 k in the yttrium-hydrogen system under
  high pressure.
\newblock {\em Nature Communications}, 12, Aug 2021.

\bibitem{maxtc}
Elliot Snider, Nathan Dasenbrock-Gammon, Raymond McBride, Mathew Debessai,
  Hiranya Vindana, Kevin Vencatasamy, Keith~V. Lawler, Ashkan Salamat, and
  Ranga~P. Dias.
\newblock Room-temperature superconductivity in a carbonaceous sulfur hydride.
\newblock {\em Nature}, 586:373--377, Octuber 2020.

\bibitem{Yinwei}
Yinwei Li, Jian Hao, Hanyu Liu, John~S. Tse, Yanchao Wang, and Yanming Ma.
\newblock Pressure-stabilized superconductive yttrium hydrides.
\newblock {\em Scientific Reports}, 5:9948, May 2015.

\bibitem{Ye2018HighHO}
Xiaoqiu Ye, Niloofar Zarifi, E.~Zurek, R.~Hoffmann, and N.~Ashcroft.
\newblock High hydrides of scandium under pressure: Potential superconductors.
\newblock {\em Journal of Physical Chemistry C}, 2018.

\bibitem{PhysRevB.76.024107}
Tetsuji Kume, Hiroyuki Ohura, Shigeo Sasaki, Hiroyasu Shimizu, Ayako Ohmura,
  Akihiko Machida, Tetsu Watanuki, Katsutoshi Aoki, and Kenichi Takemura.
\newblock High-pressure study of ${\mathrm{yh}}_{3}$ by raman and visible
  absorption spectroscopy.
\newblock {\em Phys. Rev. B}, 76:024107, Jul 2007.

\bibitem{PhysRevB.73.104105}
Ayako Ohmura, Akihiko Machida, Tetsu Watanuki, Katsutoshi Aoki, Satoshi Nakano,
  and K.~Takemura.
\newblock Infrared spectroscopic study of the band-gap closure in
  ${\mathrm{yh}}_{3}$ at high pressure.
\newblock {\em Phys. Rev. B}, 73:104105, Mar 2006.

\bibitem{PhysRevB.76.052101}
A.~Machida, A.~Ohmura, T.~Watanuki, K.~Aoki, and K.~Takemura.
\newblock Long-period stacking structures in yttrium trihydride at high
  pressure.
\newblock {\em Phys. Rev. B}, 76:052101, Aug 2007.

\bibitem{PALASYUK2005477}
T.~Palasyuk and M.~Tkacz.
\newblock Hexagonal to cubic phase transition in yh3 under high pressure.
\newblock {\em Solid State Communications}, 133(7):477--480, 2005.

\bibitem{LiYH31}
Riki Kataoka, Toru Kimura, Nobuhiko Takeichi, and Atsunori Kamegawa.
\newblock Stabilization of face-centered cubic high-pressure phase of reh3 (re
  = y, gd, dy) at ambient pressure by alkali or alkaline-earth substitution.
\newblock {\em Inorganic Chemistry}, 57(8):4686--4692, 2018.
\newblock PMID: 29620366.

\bibitem{LiYH3}
Riki Kataoka, Toru Kimura, Kouji Sakaki, Masashi Nozaki, Toshikatsu Kojima,
  Kazutaka Ikeda, Toshiya Otomo, Nobuhiko Takeichi, and Atsunori Kamegawa.
\newblock Facile synthesis of lih-stabilized face-centered-cubic yh3
  high-pressure phase by ball milling process.
\newblock {\em Inorganic Chemistry}, 58(19):13102--13107, 2019.
\newblock PMID: 31502447.

\bibitem{YH3em}
J.~Purans, A.~P. Menushenkov, S.~P. Besedin, A.~A. Ivanov, V.~S. Minkov,
  I.~Pudza, A.~Kuzmin, K.~V. Klementiev, S.~Pascarelli, O.~Mathon, A.~D. Rosa,
  T.~Irifune, and Mikhail~I. Eremets.
\newblock Local electronic structure rearrangements and strong anharmonicity in
  yh3 under pressures up to 180gpa.
\newblock {\em Nature Communications}, 12, March 2021.

\bibitem{PhysRevB.84.064132}
Tetsuji Kume, Hiroyuki Ohura, Tomoo Takeichi, Ayako Ohmura, Akihiko Machida,
  Tetsu Watanuki, Katsutoshi Aoki, Shigeo Sasaki, Hiroyasu Shimizu, and Kenichi
  Takemura.
\newblock High-pressure study of sch3: Raman, infrared, and visible absorption
  spectroscopy.
\newblock {\em Phys. Rev. B}, 84:064132, Aug 2011.

\bibitem{Kong_2013}
Bo~Kong, Zhu-Wen Zhou, De-Liang Chen, and Rong-Feng Ling-hu.
\newblock Structures and phase transitions of {ScH}3under high pressure.
\newblock {\em Chinese Physics B}, 22(5):057102, may 2013.

\bibitem{Pakornchote_2013}
T~Pakornchote, U~Pinsook, and T~Bovornratanaraks.
\newblock The hcp to fcc transformation path of scandium trihydride under high
  pressure.
\newblock {\em Journal of Physics: Condensed Matter}, 26(2):025405, dec 2013.

\bibitem{Kim2793}
Duck~Young Kim, Ralph~H. Scheicher, Ho-kwang Mao, Tae~W. Kang, and Rajeev
  Ahuja.
\newblock General trend for pressurized superconducting hydrogen-dense
  materials.
\newblock {\em Proceedings of the National Academy of Sciences},
  107(7):2793--2796, 2010.

\bibitem{Zhang_2007}
J~Y Zhang, L~J Zhang, T~Cui, Y~L Niu, Y~M Ma, Z~He, and G~T Zou.
\newblock A first-principles study of electron{\textendash}phonon coupling in
  electron-doped {LiH}.
\newblock {\em Journal of Physics: Condensed Matter}, 19(42):425218, sep 2007.

\bibitem{PhysRevB.99.214504}
M.~A. Olea-Amezcua, O.~De~la Pe\~na Seaman, and R.~Heid.
\newblock Superconductivity by doping in alkali-metal hydrides without applied
  pressure: An ab initio study.
\newblock {\em Phys. Rev. B}, 99:214504, Jun 2019.

\bibitem{Villa_Cort_s_2021}
S~Villa-Cort{\'{e}}s and O~De la~Pe{\~{n}}a-Seaman.
\newblock Electron- and hole-doping on {ScH}2 and {YH}2: effects on
  superconductivity without applied pressure.
\newblock {\em Journal of Physics: Condensed Matter}, 33(42):425401, aug 2021.

\bibitem{PhysRev.140.A1133}
W.~Kohn and L.~J. Sham.
\newblock Self-consistent equations including exchange and correlation effects.
\newblock {\em Phys. Rev.}, 140:A1133--A1138, Nov 1965.

\bibitem{vca}
Lothar Nordheim.
\newblock Zur elektronentheorie der metalle. i.
\newblock {\em Annalen der Physik}, 401(5):607--640, 1931.

\bibitem{PhysRevLett.88.127001}
H.~Rosner, A.~Kitaigorodsky, and W.~E. Pickett.
\newblock Prediction of high $t_{c}$ superconductivity in hole-doped libc.
\newblock {\em Phys. Rev. Lett.}, 88:127001, Mar 2002.

\bibitem{PhysRevLett.93.237002}
Lilia Boeri, Jens Kortus, and O.~K. Andersen.
\newblock Three-dimensional ${\mathrm{m}\mathrm{g}\mathrm{b}}_{2}$-type
  superconductivity in hole-doped diamond.
\newblock {\em Phys. Rev. Lett.}, 93:237002, Nov 2004.

\bibitem{PhysRevB.93.224513}
Yanfeng Ge, Fan Zhang, and Yugui Yao.
\newblock First-principles demonstration of superconductivity at 280 k in
  hydrogen sulfide with low phosphorus substitution.
\newblock {\em Phys. Rev. B}, 93:224513, Jun 2016.

\bibitem{PhysRevB.79.134523}
O.~De~la Pe\~na Seaman, R.~de~Coss, R.~Heid, and K.-P. Bohnen.
\newblock Effects of al and c doping on the electronic structure and phonon
  renormalization in mgb$_{2}$.
\newblock {\em Phys. Rev. B}, 79:134523, Apr 2009.

\bibitem{0953-8984-21-39-395502}
Paolo~Giannozzi et~al.
\newblock Quantum espresso: a modular and open-source software project for
  quantum simulations of materials.
\newblock {\em Journal of Physics: Condensed Matter}, 21(39):395502, 2009.

\bibitem{RevModPhys.73.515}
Stefano Baroni, Stefano de~Gironcoli, Andrea Dal~Corso, and Paolo Giannozzi.
\newblock Phonons and related crystal properties from density-functional
  perturbation theory.
\newblock {\em Rev. Mod. Phys.}, 73:515--562, Jul 2001.

\bibitem{PhysRevLett.77.3865}
John~P. Perdew, Kieron Burke, and Matthias Ernzerhof.
\newblock Generalized gradient approximation made simple.
\newblock {\em Phys. Rev. Lett.}, 77:3865--3868, Oct 1996.

\bibitem{10.2138/rmg.2010.71.3}
Stefano Baroni, Paolo Giannozzi, and Eyvaz Isaev.
\newblock {Density-Functional Perturbation Theory for Quasi-Harmonic
  Calculations}.
\newblock {\em Reviews in Mineralogy and Geochemistry}, 71(1):39--57, Jan 2010.

\bibitem{PhysRevB.6.2577}
Philip~B. Allen.
\newblock Neutron spectroscopy of superconductors.
\newblock {\em Phys. Rev. B}, 6:2577--2579, Oct 1972.

\bibitem{PhysRevB.9.4733}
Philip~B. Allen and Richard Silberglitt.
\newblock Some effects of phonon dynamics on electron lifetime, mass
  renormalization, and superconducting transition temperature.
\newblock {\em Phys. Rev. B}, 9:4733--4741, Jun 1974.

\bibitem{PhysRevB.12.905}
P.~B. Allen and R.~C. Dynes.
\newblock Transition temperature of strong-coupled superconductors reanalyzed.
\newblock {\em Phys. Rev. B}, 12:905--922, Aug 1975.

\bibitem{Eliashberg}
G.~M. Eliashberg.
\newblock Interaction between electrons and lattice vibrations in a
  superconductor.
\newblock {\em J. Exptl. Theoret. Phys.}, 38:966--976, 1960.

\bibitem{Bergmann1973}
G.~Bergmann and D.~Rainer.
\newblock The sensitivity of the transition temperature to changes in
  $\alpha$2f($\omega$).
\newblock {\em Zeitschrift f{\"u}r Physik}, 263(1):59--68, 1973.

\bibitem{VILLACORTES2018371}
S.~Villa-Cort\'es and R.~Baquero.
\newblock The thermodynamics and the inverse isotope effect of superconducting
  palladium hydride compounds under pressure.
\newblock {\em Journal of Physics and Chemistry of Solids}, 123:371 -- 377,
  2018.

\bibitem{PhysRevB.19.2540}
D.~Rainer and F.~J. Culetto.
\newblock Theory of the isotope effect in superconducting compounds: Pdd and
  ${\mathrm{mo}}_{6}$${\mathrm{se}}_{8}$.
\newblock {\em Phys. Rev. B}, 19:2540--2547, Mar 1979.

\bibitem{VILLACORTES2022110451}
S.~Villa-Cortés and O.~{De la Peña-Seaman}.
\newblock Effect of van hove singularity on the isotope effect and critical
  temperature of h3s hydride superconductor as a function of pressure.
\newblock {\em Journal of Physics and Chemistry of Solids}, 161:110451, 2022.

\bibitem{meromio}
S.~Villa-Cort\'es and R.~Baquero.
\newblock On the calculation of the inverse isotope effect in pdh(d): A
  migdal-eliashberg theory approach.
\newblock {\em Journal of Physics and Chemistry of Solids}, 119:80 -- 84, 2018.

\bibitem{doi:10.1063/1.4714549}
X.~Q. Zeng, L.~F. Cheng, J.~X. Zou, W.~J. Ding, H.~Y. Tian, and C.~Buckley.
\newblock Influence of 3d transition metals on the stability and electronic
  structure of mgh2.
\newblock {\em Journal of Applied Physics}, 111(9):093720, 2012.

\bibitem{PhysRevB.69.094205}
Y.~Song, Z.~X. Guo, and R.~Yang.
\newblock Influence of selected alloying elements on the stability of magnesium
  dihydride for hydrogen storage applications: A first-principles
  investigation.
\newblock {\em Phys. Rev. B}, 69:094205, Mar 2004.

\bibitem{MACHIDA2006436}
A.~Machida, A.~Ohmura, T.~Watanuki, T.~Ikeda, K.~Aoki, S.~Nakano, and
  K.~Takemura.
\newblock X-ray diffraction investigation of the hexagonal–fcc structural
  transition in yttrium trihydride under hydrostatic pressure.
\newblock {\em Solid State Communications}, 138(9):436--440, 2006.

\bibitem{Birch}
F.~D. Murnaghan.
\newblock Finite deformations of an elastic solid.
\newblock {\em American Journal of Mathematics}, 59(2):235--260, 1937.

\end{thebibliography}

\end{document}